\input{epsf}
\documentstyle[prd,aps]{revtex}

\begin{document}
\draft
\title{Gravitational waves from pulsating stars: Evolving the perturbation 
equations for a relativistic star}

\author{
Gabrielle Allen$^{1,2}$,
Nils Andersson$^{3}$,
Kostas D. Kokkotas $^{1,4}$
Bernard F. Schutz $^{1,2}$}

\address{$^1$ Max Planck Institute for Gravitational Physics,
The Albert Einstein Institute,
D-14473 Potsdam, Germany}
\address{$^2$ Department of Physics and Astronomy,
University of Wales College of Cardiff,
Cardiff CF2 3YB, United Kingdom}  
\address{$^3$ Department of Physics,  Washington University, 
St Louis MO 63130, USA}
\address{$^4$ Department of Physics, Aristotle University of Thessaloniki,
Thessaloniki 54006, Greece}


\maketitle

\begin{abstract}
\widetext
 We consider the perturbations of a relativistic star as an initial-value problem. 
Having discussed the formulation of the problem (the perturbation 
equations and the appropriate boundary conditions
at the centre and the surface of the star) in detail we evolve the equations
numerically from several different sets of initial data. In all the considered
cases we find that the resulting gravitational
waves carry the signature of several of the star's pulsation modes. Typically, the fluid $f$-mode, the first two $p$-modes and the slowest damped gravitational $w$-mode are present in the signal. This indicates that
the pulsation modes may be an interesting source for detectable gravitational waves from colliding neutron stars or supernovae. We also survey the literature
and find several indications of mode presence in numerical simulations of
rotating core collapse and coalescing neutron stars. 
If such mode-signals 
can be detected by future gravitational-wave antennae one can hope to
infer detailed information about neutron stars. 
Since a 
perturbation evolution should
adequately describe the late time behaviour of a dynamically
excited neutron
star, the present work can also be used as a bench-mark test for
future fully nonlinear simulations.

\end{abstract}



\section{Introduction}

It is well known that a neutron star has a rich pulsation spectrum
  \cite{mcdermott83,mcdermott88,gau95,andersson96a}.
 We expect the stellar pulsation modes to be excited in many neutron
  star processes (ranging from core quakes to the formation of a neutron
star  through gravitational collapse). 
A signal carrying the signature of  these modes
could,
  if detected by future gravitational-wave antennaes, 
  provide useful information about the star. 
The question is whether the various stellar pulsation modes can be
 dynamically excited to a level that makes them astrophysically
relevant. This question is especially interesting because of the
existence of ``spacetime'' modes  \cite{kokkotas92},
which have no analogue in the Newtonian theory of stellar pulsation. 
It has been shown \cite{andersson96d} that these modes, together with
the normal fluid pulsation modes, can provide valuable information about
the mass, size, and equation of state of neutron stars. 
Much of the initial deformation of spacetime in e.g. a supernova collapse 
could conceivably be released
 through spacetime pulsation modes, yet most studies of neutron star
 formation and pulsation have not treated them because they have been done
 within the Newtonian or post-Newtonian approximations.

With the present paper we take the first steps towards an answer to
the
 question of mode-excitation. 
We describe numerical evolutions of the equations that 
 describe a perturbed relativistic star for various sets 
 of initial data.
Evolutions of the perturbation equations for black holes have provided
 interesting information (see 
  \cite{nbook} 
 for references),
and it seems likely that this approach should prove equally instructive for
stars. Our main focus is on the evolution problem itself:
We formulate the problem and discuss all relevant equations in some
detail. 
We also present results from two test problems, which indicate
 that for a range of initial conditions the energy emitted is shared
 broadly among the pulsation and spacetime modes. This argues strongly, if in 
 a preliminary way, that neutron-star dynamical calculations done within the
 Newtonian and post-Newtonian approximations may seriously 
 underestimate the amount and the spectral character of the emitted 
 gravitational radiation.

The plan of the paper is as follows. In Section II we formulate the
linear perturbation problems for a neutron star in general relativity
as an initial-value problem, with special attention to the boundary 
conditions and the treatment of the center of the star. In Section III
we solve the initial-value problem numerically for two sets of initial 
conditions, which represent in some sense the extremes of the balance between 
exciting fluid modes on the one hand and spacetime modes on the other. 
The spectrum of the gravitational waves generated from these initial data show
clearly the signatures of both the fluid modes and the spacetime modes. 
Importantly, in both sets the excitation of the spacetime modes is
appreciable. In Section IV we discuss the significance of the results, and
our plans for future work to determine a more realistic initial-data
set to use for such calculations. We begin by providing, in the
remainder of Section I, a more detailed motivation for this work in
the context of numerical relativity and gravitational wave detection,
and a brief background on relativistic stellar pulsation theory with special
focus on recent developments.

\subsection{Gravitational-wave astronomy and numerical relativity}

With the building of several large-scale laser interferometers for
 gravitational-wave detection well under way 
 \cite{ligo,virgo} 
``gravitational-wave astronomy'' may be established around the turn of 
the century.  
But many fundamental problems remain to be solved before this 
 goal can be reached.
For the theorist, the most pressing problems concern modeling 
 the processes which generate gravitational waves.
Only by comparing observational data to such models can we hope to infer 
 detailed information about the various gravitational-wave sources. 

Promising sources for detectable gravitational waves
 are the coalescence of compact binaries
 (that is, the spiraling collision of a binary system 
  containing neutron stars and/or black holes) 
 and supernovae collapse 
  \cite{lhouches96}.
A complete description of these strong field, highly asymmetric 
 processes requires fully relativistic 3-dimensional numerical
simulations. There is currently a ``Grand Challenge'' project 
 underway in the USA to simulate the 3-dimensional spiraling 
 coalescence of black holes 
  \cite{gc95}. Numerical simulations of the head-on 
(axisymmetric) collision of 
 two black holes have already been performed 
  \cite{anninos93}, 
 providing preliminary estimates for the gravitational-wave 
 efficiency of this process.
These calculations indicate that the quasinormal modes of the
 final black hole dominate the emerging radiation.  

Simulations of processes involving neutron stars have
 (with a few exceptions) so far used Newtonian gravity\cite{bonnazolla},
see the discussion in Section IIIC. 
This is mainly due to the difficulties of consistently treating 
 both the matter and the gravitational field within the framework
 of the standard ADM formalism.
In these Newtonian simulations the gravitational radiation is calculated 
 using the quadrupole formula.
That is, gravitational waves originate solely from stellar fluid motion 
 --- there is no contribution from the dynamics 
 of the gravitational field itself. 
However, the dynamics of the gravitational field 
should also contribute to the emerging gravitational radiation. 
This is clear from the black hole case, 
 in which the quasinormal-mode oscillations are entirely due to 
 the dynamical spacetime 
  \cite{nbook}.
At an intuitive level one would  expect similar features to  
 exist for stars, although they are probably less 
 dominant since a neutron star is less 
 relativistic than a black hole. 
As we describe below this is, indeed, the case. 
There is a set of pulsation modes of a relativistic star that 
 can be directly associated with the spacetime curvature.
But it is not yet known how important the spacetime contribution 
 to the gravitational-wave emission will be in a dynamical scenario.

There is growing optimism in the numerical relativity
 community over the feasibility of simulating neutron stars within
general relativity.  
This is mainly due to the recent development of ``hyperbolic
 formalisms''  for vacuum relativity into which relativistic
hydrodynamics 
 fits  naturally. 
There is now another ``Grand Challenge'' project in the USA 
 designated to simulate binary neutron star coalescence, and there are
 plans in Europe to simulate supernovae collapse --- both fully relativistically. 
While such projects are ultimately the only way to provide 
 reliable gravitational-wave estimates, 
the first results will not be available for some years. 

In the meantime, we must resort to approximate methods such as 
 post-Newtonian expansions and perturbation theory. 
These methods can provide valuable information 
 about the early and late time behaviour of the system. 
The present paper describes a project to investigate the excitation of 
 the stellar pulsation modes using perturbation theory. 
We recall that non-linear numerical calculations for 
 black holes have shown that the quasinormal modes tend to be clearly
present in the radiation.
Furthermore, comparisons with 
 non-linear calculations have indicated that
 the more tractable linear approximations can provide accurate 
 waveforms for a wide range of initial data. A particularly 
 interesting example of this is the ``close-limit'' 
 approximation for colliding black holes \cite{price94}. 

It seems likely that perturbation 
calculations will also be a useful tool for problems 
involving relativistic stars. 
  The relative simplicity of the perturbation approach compared 
  to the full non-linear calculations makes its application and
  interpretation easier. We can 
  hope to learn some of the relevant physics by pursuing the 
perturbation approach. 
In the near future, 
 while non-linear codes are still being developed, 
 approximate studies can lead the way.
A further benefit from this work is its application as a test for 
 the full non-linear codes for simulations of stellar processes.
One can immediately identify two ways in which results of a fully
 relativistic numerical simulation should agree 
 with a perturbation result, and
in the case of black holes these tests are well
 established:
i) Extract  asymmetric perturbations of the final spacetime
   from the numerical waveforms and compare these perturbations
   to the anticipated quasinormal modes, cf. \cite{anninos93}.  
ii) Apply the fully non-linear code to ``linear initial data''
    and compare the results with an independent perturbation 
    calculation \cite{abrahams}.

\subsection{Pulsating stars}

The study of stellar pulsation in a general relativistic context
 has a considerable history, dating back to the fundamental 
 work of Thorne and his colleagues in the late 1960's
  \cite{thorne67} (for a recent review see 
  \cite{kreview}). 
Originally,  the stellar pulsation problem was approached from 
a ``Newtonian'' viewpoint. 
In this picture, motion in the stellar fluid generates gravitational 
 waves which carry energy away from the system. The characteristic 
 frequencies of the various pulsation modes of the star thus 
 become complex valued. 
Due to the weak coupling between matter and gravitational waves the
 damping rate of a typical pulsation mode is very low   
 \cite{lindblom83}.

This ``Newtonian'' picture is limited in that the dynamical properties 
 of the spacetime itself are neglected.
That a dynamic spacetime can add new features to the pulsation
problem can be understood in terms of a simple, but instructive, 
model problem 
due to Kokkotas and Schutz  \cite{kokkotas86}. 
There are pulsation modes directly 
 associated with spacetime itself. 
 These new modes have been termed ``gravitational-wave modes'' 
 (or $w$-modes)
  \cite{kokkotas92}. 
They have relatively high oscillation frequencies 
 (6-14 kHz for typical neutron stars) 
 and barely excite any fluid motion. 
They are also rapidly damped, with a typical lifetime of a fraction 
  of a  millisecond. Our
understanding of the $w$-modes has improved with a body of recent work
  \cite{andersson96a,andersson96b,kojimaprs,andersson96c}. 
A present conjecture \cite{andersson96a} is that three 
(more or less
distinct) classes of gravitational-wave modes exist: 
   i) curvature modes; that arise when gravitational waves are trapped 
      in the bowl of curvature associated with the star, 
  ii) interface modes; that are closely linked to the stellar surface,
and 
 iii) trapped modes; that arise when the star is so compact 
      ($R<3M$) 
      that the peak of the exterior curvature potential
      (familiar from the black-hole problem) is unveiled.

The existence of pulsation modes that are directly associated with the
 spacetime itself is interesting from a theoretical point of view,  
but it is necessary to establish if these modes are of 
 astrophysical significance \cite{andersson96d}.
One can argue that the gravitational-wave modes 
could be relevant in many scenarios. 
 Consider, for example, gravitational collapse to a neutron star, 
or the coalescence of two neutron stars. In both cases will there
be changes in the deformation of spacetime, which could potentially lead
to considerable amounts of energy being radiated  through the $w$-modes.
Detailed calculations are
needed to provide quantitative information.
In essence, there are two important questions that must addressed: i)
Will the 
gravitational-wave modes be excited in processes such as supernova
collapse or neutron 
 star coalescence? ii) What is the possibility of observing such 
 modes in data from the new generation of gravitational-wave detectors, 
and what physical
 information could such observations provide?

As an initial attempt to answer the first question, we studied
scattering of
wave-packets by a uniform density star \cite{andersson96d}. 
This is not a problem of great astrophysical importance (it is 
difficult to imagine a situation where the impinging gravitational wave has sufficient magnitude to make the scattered wave observable) but 
the results were
 nevertheless encouraging. The resultant gravitational
 waves show the clear signature of the $w$-modes 
for neutron star sized objects ($R/M\approx 5$). 
Similar results
 have since been obtained for a particle falling in the stellar
spacetime \cite{valeria}.
The present paper describes more detailed work along these lines: 
We evolve the perturbation equations from general initial data. 
Most importantly, our analysis is not restricted to an initial 
perturbation with compact support in the
vacuum outside the star. To construct our stellar model we use a 
polytropic equation of state.
Contrary to the simple uniform density model, polytropes allow
several fluid pulsation modes to exist (the single $f$-mode and
an infinite sequence of $p$-modes).  Our aim
 is to demonstrate that our expectations regarding the excitation of the
stellar pulsation modes are reasonable, and that these modes
 should be considered as a source for detectable gravitational waves.

The second question --- the possibility of extracting information from a
detected
pulsation mode --- has been
considered by Andersson and Kokkotas 
  \cite{andersson96d}.
Their preliminary investigation suggests that one may be able to
 extract \underline{both} the mass and the radius of a 
 perturbed star using observed mode-signals. 
Such information can help put constraints on the nuclear equation of
state.

\section{Stellar Perturbations as an Initial Value Problem}

In this section we describe the 
stellar perturbation problem. We will introduce 
all the necessary equations but not discuss their
origin in great detail. For further details we refer the reader
to previous work\cite{thorne67,lindblom83,detweiler85,ipser91,kojima92}.
 It is relevant to point out that most of the existing 
work on perturbations of stars has been performed in the 
frequency domain (after Fourier decomposition of the
various perturbed quantities). This is the natural approach as
long as one is mainly interested in the spectral properties of a star.
Kind, Ehlers and Schmidt \cite{kind93} appear to be the only authors that
have considered relativistic stellar perturbations as an initial value problem.
(Perturbations of a time-dependent geometry were 
considered by Seidel  et al. \cite{seidel87,seidel88,seidel90} in the context of gravitational collapse). 
The motivation of Kind et al. was to show that initial Cauchy 
data with the appropriate junction conditions at the stellar surface
determines a unique solution to the time-dependent equations. 
Thus, Kind {\em et al.} address important mathematical questions
for the stellar evolution problem.
The motivation of the present investigation is rather different, but
our formulation of the problem is still inspired
by \cite{kind93}. 

\subsection{The stellar model}

A static spherically symmetric stellar model can be described by the metric
\begin{equation}
ds^2 = -e^\nu dt^2 + e^\lambda dr^2 + 
r^2 \left( d\theta^2 +\sin^2 \theta d\varphi^2 \right) \ ,
\end{equation}
where the metric coefficients $\nu$ and $\lambda$ are 
functions of $r$ only. Specifically,
\begin{equation}
e^\lambda = \left( 1 - {2 m(r) \over r} \right)^{-1} \ ,
\end{equation}
and the ``mass inside radius $r$'' is represented by
\begin{equation}
m(r) = 4\pi \int_0^r \rho r^2 dr  \ .
\end{equation}
This means that the total mass of the star is $M=m(R)$, with 
$R$ being the
stars radius. (We use geometric units $G=c=1$ throughout this paper.)

A star in hydrostatic equilibrium is governed by the TOV equations:
To determine a stellar model we must solve
\begin{equation}
{dp \over dr} = - {\rho + p \over 2} {d\nu \over dr} \ ,
\end{equation}
where
\begin{equation}
{d\nu \over dr} = {2 e^\lambda(m + 4\pi p r^3) \over r^2} \ .
\end{equation}
These equations do, of course, require an equation of state
$p=p(\rho)$ as input. In this paper we consider (for reasons of 
simplicity) only polytropic equations of state
\begin{equation}
p = \kappa \rho^\gamma \ .
\label{polyt}\end{equation}
In particular, we present numerical results for a model with
$\kappa=100 \mbox{ km}^2$ and $\gamma=2$. 
We have performed calculations using different parameters but,
since the results were similar to those discussed in section IIIB, we
will only discuss one stellar model in detail. The specific model we
have chosen has a central density $\rho_c = 3\times 10^{15} $ g/cm$^3$. 
The corresponding radius and mass are
$R= 8.86$ km  and $M= 1.87 \mbox{ km }\approx 1.2 M_\odot$ , respectively. 
The pulsation properties of this stellar model have already been 
investigated in detail previously \cite{kokkotas92}.

\subsection{The perturbation problem}

A general description of the stellar perturbation
problem is as follows:
We want to solve the perturbed Einstein equations 
\begin{equation}
\delta G_{\mu \nu} = 8\pi \delta T_{\mu\nu} \ .
\end{equation}
The small amplitude motion that a perturbation induces in the
stellar fluid is described by displacements 
$\xi^r$, $\xi^\theta$ and $\xi^\varphi$. The fluid displacement
affects also
the pressure and the density of the fluid. To describe these we use
the Eulerian variations
$\delta p$ and $\delta \rho$.
 
As is familiar from 
black-hole problems\cite{nbook,chandrabook}, the perturbed metric can be split
into two classes: axial and polar (often called odd and even parity)
perturbations. 
That is, the metric can be written
\begin{equation}
g_{\mu \nu} = g_{\mu \nu}^{\rm background} + h_{\mu \nu}^{\rm polar} +  
h_{\mu \nu}^{\rm axial} \ .
\end{equation}
If we work
in Regge-Wheeler gauge (which has been used for most of the previous 
studies of the stellar problem)
the axial perturbations
are described by four variables linked to the metric coefficients 
$h_{t\theta}$, $h_{t\varphi}$, $h_{r\varphi}$ and $h_{r\theta}$.
These perturbations induce a differential rotation of the star where
the only non-zero component of the fluid-displacement vector is 
$\xi^\varphi$.
The polar metric perturbations correspond to six variables related to
the metric coefficients $h_{tt} $, $h_{tr}$, $h_{rr} $ 
$h_{\theta \theta}$, $h_{\theta \varphi}$ and $h_{\varphi \varphi}$
together with the fluid displacements
$\xi^r$ and $\xi^\theta$.
Exactly as for black holes, it turns out that
the equations that govern the axial and the polar quantities 
decouple in the case of a nonrotating star (this will not
be the case if the star is rotating\cite{kojima92,chandra91a}). 
We thus have two problems that can be approached separately.

In this paper we only consider the polar problem.
There are several reasons for this. 
We have recently considered the initial-value
problem for axial perturbations of a uniform density star
\cite{andersson96d}.
Because an axial perturbation does
not induce pulsations in the stellar fluid \cite{thorne67} we do not expect 
the results
for a more realistic stellar model to be dramatically 
different from those for
uniform density. Basically, an axial perturbation can only
result in the excitation of $w$-modes, and  as long as the two stars
have similar compactness ($R/M$) the corresponding 
frequencies are quite similar for a uniform density
star and a more realistic model \cite{andersson96a}.
Hence, we focus on the
polar problem here. 

The polar problem corresponds to seven perturbed field equations
and  three equations of motion for four metric perturbations and 
four fluid variables\cite{thorne67,kojima92,kind93}. 
At first this would seem to be an
overdetermined problem, but this is not the case. 
Three of our original equations (the equations of motion, say) 
are void of new 
information
because of the Bianchi identities. 
Moreover, we can use 
\begin{equation}
\delta \rho = {1 \over C_s^2} \delta p \ ,
\end{equation}
where $C_s$ is the acoustic wave-speed in the stellar fluid.
We are thus left with seven equations for seven unknown variables. 

\subsection{Evolution equations}

There are many possible formulations of the perturbation 
problem, even within the Regge-Wheeler gauge, and it
 is difficult to predict if a specific one may be
advantageous from the point of view of numerical evolutions.
Here we have chosen a formulation based on the
expected physics of the system. That is, we consider a set of
wave-equations describing gravitational waves coupled to acoustic waves in the
stellar fluid. Furthermore, we have cast the ``spacetime'' 
equations into a form which resembles
the more familiar equations for black-hole perturbations. This facilitates
a comparison between the two problems.

The perturbed Einstein equations and the fluid equations of motion 
can be manipulated into a set of evolution equations for a reduced 
number of variables.
The remaining spacetime and fluid perturbations are
determined (at all times) by a number of constraint equations.
 Here we evolve two 
spacetime variables $F(t,r)$ and $S(t,r)$ which are related to the 
metric perturbations by 
\begin{eqnarray}
&&h_{\theta\theta}(t,r)  =  r F(t,r) \ , \\
&&h_{tt}(t,r)-\frac{e^\nu}{r^2} 
  h_{\theta\theta}(t,r)  =  r S(t,r)  \ .
\end{eqnarray}
These functions are related to the standard Regge-Wheeler functions $K(t,r)$
and $H_0(t,r)$ (see e.g. \cite{thorne67} or \cite{kojima92}) 
by $F = r K$, $S = e^\nu(H_0-K)/r$.
We also evolve a fluid function (the perturbed relativistic enthalpy)
\begin{equation}
H(t,r) = {{\delta p}\over{\rho +p}} \ ,
\end{equation}
which clearly is defined only inside the star.
In these definitions of $F,S$ and $H$ we have suppressed 
the angular dependence \cite{angular}.

The above variables are not obvious, 
but we have good reasons for introducing
them. Let us first consider the function $S$. When we introduce
this function, and also use the standard definition of the ``tortoise
coordinate''
\begin{equation}
 {{\partial}\over {\partial r_*}} =
 e^{(\nu-\lambda)/ 2}{{\partial}\over {\partial r}}  \ ,
\label{tort}\end{equation}
we find that \cite{origins}
\begin{eqnarray} 
-{{\partial ^2 S}\over {\partial t^2}} 
         &+&  {{\partial ^2 S}\over {\partial r_*^2}} 
         +  {{2 e^\nu}\over r^3}  
             \left[ 2 \pi r^3 (\rho+3p) + m - (n+1)r \right] S 
         \nonumber \\ &=& 
          -{{4 e^{2\nu}}\over r^5} \left[ 
    {{(m+4\pi p r^3)^2} \over {(r-2m)}} + 4\pi \rho r^3 - 3m \right] F \ , 
\label{Swave} 
\end{eqnarray} 
where $2n = (l-1)(l+2)$.
We thus have a simple wave-equation with no explicit dependency
on the perturbations of the fluid. For the other ``metric'' variable
$F$ we find \cite{origins}
\begin{eqnarray} 
-{{\partial ^2 F}\over {\partial t^2}} 
     &+& {{\partial ^2 F}\over {\partial r_*^2}}  
      +  {{2 e^\nu}\over r^3} \left[2 \pi r^3 (3 \rho+ p) + m - (n+1) r \right] F 
    \nonumber \\ &=& 
         - 2\left[ 4 \pi r^2 (p+\rho) - e^{-\lambda} \right] S 
         +  8 \pi (\rho+p) {r e^{\nu}} 
            \left(1- {1\over C_s^2} \right) H \ .
\label{Gwave} 
\end{eqnarray} 
This equation is also quite simple, but here the coupling to the
fluid variable $H$ is apparent.
Both of our wave equations are valid inside the star as well as in the
exterior vacuum. In the exterior the mass $m$ is, obviously, 
the total mass $M$. 
 Our reason for introducing
$S$ and $F$ should now be apparent: In terms of these variables the
perturbation equations become similar to the wave-equation that
governs a perturbed Schwarzschild black hole \cite{nbook}, or the
equation for axial perturbations of a star\cite{chandra91b}.
That the polar equations can be cast
in a similar form has not been shown previously. 

 The equation that governs the fluid variable $H$ is, however, still somewhat
messy. After some algebra we find that \cite{origins}
\begin{eqnarray} 
- {1\over {C_s^2}}{{\partial ^2 H}\over {\partial t^2}} 
      &+&  {{\partial ^2 H}\over {\partial r_*^2}} 
      + {e^{(\nu+\lambda)/2} \over r^2} 
      \left[(m + 4\pi p r^3)\left(1-{1\over {C_s^2}}\right) + 2 (r-2m) \right] 
        {{\partial H}\over {\partial r_*}} 
        \nonumber \\ 
  &+& {{2 e^\nu}\over r^2} 
     \left[ 2 \pi r^2 (\rho+p)\left(3 + {1\over {C_s^2}} \right) 
     - (n+1) \right] H 
     \nonumber \\ 
     &=& (m+4 \pi p r^3)\left(1-{1\over {C_s^2}}\right) 
     {{e^{(\lambda-\nu)/2}}\over {2 r}} 
     \left[ 
         {{e^\nu}\over r^2}{{\partial F}\over {\partial r_*}} 
        -  {{\partial S}\over {\partial r_*}} 
     \right] 
     \nonumber \\ &+& 
     \left[ 
      {{(m+4\pi p r^3)^2}\over{r^2(r-2m)}}\left(1+{1\over {C_s^2}}\right) - 
      {{ m+4\pi p r^3   }\over{2 r^2}    }\left(1-{1\over {C_s^2}}\right) 
      -4\pi r (3p+\rho) 
      \right] S 
      \nonumber \\ &+& 
     {{e^\nu}\over r^2}\left[ 
     {{2(m+4\pi p r^3)^2}\over{r^2(r-2m)}}{1\over {C_s^2}} - 
      {{ m+4\pi p r^3   }\over{2 r^2} }\left(1-{1\over {C_s^2}} \right) 
      -4\pi r (3p+\rho) 
      \right] F  \ .
\label{Hwave} 
\end{eqnarray}

Finally we have the Hamiltonian constraint \cite{origins}
\begin{eqnarray} 
{\cal H} := {{\partial ^2 F}\over {\partial r_*^2}} &-& 
 {e^{(\nu+\lambda)/2} \over r^2} \left( m + 4 \pi r^3 p \right) 
{{\partial F}\over {\partial r_*}} 
+ {{e^\nu}\over r^3} \left[ 12\pi r^3 \rho - m - 2(n+1)r \right] F 
\nonumber \\ 
&-& r e^{-(\nu+\lambda)/2}{{\partial S}\over {\partial r_*}} + 
\left[8\pi r^2(\rho+p) -(n+3) + {4m \over r} \right] S 
+ {{8 \pi r}\over {C_s^2}} e^\nu (\rho+p) H = 0\ ,
\label{Hamilton}\end{eqnarray}
 which
must be satisfied by the initial data and also throughout an evolution. 
We will discuss this constraint in more 
detail in section IIIA.

Here it could be worthwhile to note that it is possible to find a 
new ``fluid'' variable
such that the corresponding wave equation [that would replace (\ref{Hwave})]
contains no first derivatives. However, this extra step
is not convenient because it involves the solution of a differential
equation involving $C_s^2$. 

It is also meaningful to comment on the number of equations that we use.
It is well known that the interior problem for polar perturbations 
can be reduced to two 
coupled wave equations \cite{ipser91,kojima92}. 
It would be 
easy to use (\ref{Hamilton}) to replace $H$ in 
(\ref{Gwave}) (as was done by Ipser and Price \cite{ipser91} and 
Kojima\cite{kojima92}). If we do this we get
\begin{eqnarray}
&&  - { 1 \over C_s^2} {\partial^2 F \over \partial t^2}
+ {\partial^2 F \over \partial r_*^2} 
  - \left( 1 - { 1\over C_s^2} \right) {e^{(\nu+\lambda)/2} \over r^2} 
(m+4 \pi p r^3) {\partial F \over \partial r_*} \nonumber \\
&+&{e^\nu \over r^3} \left[ 4\pi r^3 \left(3\rho + {p \over C_s^2} \right) -
m\left(1 - {3\over C_s^2} \right) - 2(n+1)r \right] F \nonumber \\
\\
& = & \left( 1 - { 1\over C_s^2} \right) r e^{-(\nu+\lambda)/2}
{\partial S \over \partial r_*}
  \nonumber \\
&+& \left[ 2e^{-\lambda} + \left( 1 - { 1\over C_s^2} \right) (n+1)
- 8\pi(p+\rho)r^2 \right] S
\label{Ginside}\end{eqnarray}
However, (\ref{Ginside}) is rather ``unphysical''
because a metric variable  ($F$) plays the role of the fluid. 
It is also useful to keep one fluid variable explicit.
With the equations for $S$, $F$, and $H$ above we can 
easily monitor the fluid motion during the
evolution. Moreover,
Kind {\em et al.}
\cite{kind93}  have shown that our problem is
well posed (that a unique solutions exists).

Let us now turn to the exterior problem. In the vacuum outside the star
the problem is identical to that for a perturbed Schwarzschild black hole.
Thus, the equations simplify considerably, and 
we only need to consider the two equations for $S$ and $F$ 
(note that the two effective potentials are equal in the exterior vacuum).
Furthermore, we know from the studies of perturbed black holes that
we can reduce the problem even further. We only need to consider
one single homogeneous wave equation -- the
Zerilli equation \cite{nbook}. If we define\cite{seidel90,moncrief74}
\begin{equation}
{ (n+1) (nr+3M) \over r^3} 
Z = - {1\over r} {\partial F \over \partial r_\ast} + { (n+2) r - M \over r^3} F 
+ S \ , 
\label{Zdef}\end{equation}
we find that $Z$ evolves according to
\begin{equation}
{\partial^2 Z \over \partial t^2} - {\partial^2 Z \over \partial r_\ast^2} + V_Z(r) Z = 0
\end{equation}
where the effective potential is given by
\begin{equation}
V_Z(r) = {2 e^\nu \over (nr+3M)^2 r^3} \left[ n^2(n+1)r^3 + 3n^2Mr^2 + 9nM^2 r + 9M^3 \right]
\end{equation}

This equation has the advantage that the function $Z$ is gauge-invariant
and unconstrained \cite{seidel90,moncrief74}.
But from the point of view of a numerical evolution it may be 
preferable to solve the coupled system for $S$ and $F$ -- the variables
that are used inside the star -- rather than
switching to the Zerilli equation in the exterior. Furthermore, it
turns out that it is trivial to obtain gauge-invariant (but
constrained) quantities
from $S$ and $F$. If we use $q_1$ and $q_2$ to denote the gauge invariant
variables that were first introduced by Seidel \cite{seidel90},
we find that
\begin{eqnarray}
q_1 &=& H_0 = {F\over r} + r e^{-\nu} S \ , \\
q_2 &=& K = {F\over r} \ .
\label{ginvar}\end{eqnarray}
These equations prescribe how the gauge invariant quantities can be generated
from the specific functions that we solve for in Regge-Wheeler gauge. 
It is straightforward
to generate $q_1$ and $q_2$ once we know $S$ and $F$ (or the standard
Regge-Wheeler functions $H_0$ and $K$). Furthermore, for the
fluid variable one finds that $\delta \rho$ in Regge-Wheeler gauge 
corresponds immediately to the gauge-invariant variable $\Delta \eta$ 
that was used by Seidel \cite{seidel90}.

\subsection{Behaviour at the centre and the surface of the star}

When evolving the perturbation equations one must give
special consideration to the behaviour of the variables close
to the centre and the surface of the star. 

To infer the behaviour at the centre is relatively easy because
we know that, for a physical solution, all the perturbed variables should be 
regular at $r=0$. Working in the Fourier domain and 
expanding all variables in powers of $r$ (as was done by
Lindblom and Detweiler \cite{lindblom83}) we can infer that 
\begin{equation}
\left. 
\begin{array}{lll}  
 & F  \sim  r^{\ell+1} \\
 & S  \sim  r^{\ell+1} \\
 & H  \sim  r^{\ell} 
\end{array} \right\} \mbox{ as } r\to 0 \ .
\label{centre}\end{equation} 

We also need to deduce the behaviour of the 
various functions at the surface of the
star. 
We must implement the usual junction conditions \cite{mtw} for the spacetime 
functions at $r=R$, and also impose a boundary condition on the fluid variable $H$.
(It is, of course, also necesssary to
specify the behaviour of $S$ and $F$ as $r\to \infty$. But this is
easy since the two functions should correspond to purely outgoing 
waves far away from the star.)

The  surface of the star is formally defined by the vanishing of the 
Lagrangian variation in the pressure $\Delta p$. That is, at the surface
we have
\begin{equation}
\Delta p = \delta p + \xi^r {dp \over dr} = 0 \ .
\end{equation}
This immediately leads to  $\delta p \sim \rho \xi^r$ as $r\to R$, or
\begin{equation}
H \sim \xi^r \ , \quad \mbox{as } r\to R \ .
\label{Hbeh}\end{equation}
In principle, this constitutes a boundary condition for $H$ at the 
surface of the star, but in practice this result is not very helpful. 
First of all we are not explicitly calculating $\xi^r$. Moreover, it 
is known from 
Newtonian pulsation theory that
 $\xi^r$ is not constrained at 
the surface of the star (it need only be non-singular). 
 Thus,
the vanishing of $\Delta p$ as $r\to R$ does not lead to a useful 
boundary condition for the fluid variable $H$. 
However, it is known that a proper solution to Einstein's equations
will enforce the vanishing of $\Delta p$ at the surface of the star. Any solution that follows from our system of equations should therefore be
acceptable. 

This means that the behaviour of $H$ at $r=R$ can be inferred from (\ref{Hwave}).
However, because the sound speed $C_s$ vanishes as $r\to R$  
it is not convenient to use
equation (\ref{Hwave}) close to the surface. Instead
we infer the behaviour of $H$ at the stellar surface by taking the limit 
$C_s\to 0$ in (\ref{Hwave}). Keeping only terms of order $1/C_s^2$  we get
\begin{equation} 
{\ddot H} +{M\over R^2}{{\partial H}\over {\partial r_*}} 
- {M\over{2R^3}}{{\partial F}\over {\partial r_*}} 
+ {M\over {2(R-2M)}}{{\partial S}\over {\partial r_*}} 
+ {{M(R+2M)}\over{2R^5}}F
+ {M\over{2R(R-2M)}}S=0  \ ,  
\label{Hsurf}\end{equation}  
at $r=R$.
We use this equation to evolve $H$ at the surface of the star.

Let us now consider the behaviour of the two ``metric'' variables at
$r=R$. 
From the junction conditions (the continuity of the first and second 
fundamental forms across the surface of the star \cite{kind93}) 
we find that $F$, $S$, and $\partial S/\partial r_\ast$  are continuous
at $r=R$.
That is, if we let 
subscripts $i$ and $e$ represent functions obtained for the
interior and the exterior, respectively, we have
\begin{equation}
 F_i=F_e \ , \qquad
 S_i=S_e \ , \qquad 
{\partial S_i\over \partial r_\ast}  = {\partial S_e\over \partial r_\ast} \ .
\end{equation}
It is also apparent from (\ref{Swave}) that these conditions imply
\begin{equation}
{\partial^2 S_i\over \partial r_\ast^2} = 
{\partial^2 S_e\over \partial r_\ast^2} \ ,
\end{equation}
as long as $\rho\to 0 $ as $r\to R$.
This means that 
all terms in (\ref{Swave}) are well defined at the boundary,
and this equation can consequently be used to evolve $S$ there.

For the function $F$ the situation is somewhat different. First we 
notice that we can derive an equation similar to (\ref{Hsurf}) also
for $F$ ({\em i.e.} an equation with no second derivative with respect to 
$r_\ast$). 
This equation is
\begin{equation} 
{\ddot F} -{M\over R^2}{{\partial F}\over {\partial r_*}} 
- R{{\partial S}\over {\partial r_*}} 
- {3M \over R^3}\left( 1 - {2M\over R} \right)F 
- (n+1)S =0 \  , 
\label{Gsurf}\end{equation}
at $r=R$.
Using this equation, and the already established continuity of 
$\partial S/\partial r_\ast$, we can infer that
\begin{equation}
{\partial F_i\over \partial r_\ast}  = 
{\partial F_e\over \partial r_\ast} \ .  
\end{equation}
Thus,  $\partial F / \partial r_\ast$ is continuous across the surface of
the star. 

Finally, it follows from the wave equation (\ref{Gwave}) that
 \begin{equation}
{\partial^2 F_i\over \partial r_\ast^2} - 
{\partial^2 F_e\over \partial r_\ast^2}  = 
8\pi R e^{\nu(R)} H(R) \lim_{r\to R} {p+\rho \over
C_s^2} \to 8\pi R e^{\nu(R)} H(R) {\rho^{2-\gamma} \over 
\kappa \gamma}\mbox{ as } r\to R \ ,
\label{2der}\end{equation}
where $\kappa$ and $\gamma$ are the constant and index in the 
polytropic equation of
state, respectively.
Thus we see that in the case that we consider ($\gamma=2$) the 
right hand side of (\ref{2der}) approaches a constant as $r\to R$. 
Consequently, the second derivative of $F$ will be discontinuous
across the surface of the star.

\section{Two model problems}

In the previous section we discussed the equations that govern
perturbations of a relativistic
star. We now want to evolve these equations from a given set of initial 
data. The ultimate purpose of this exercise is to use the perturbation 
approach to infer what one should expect in a problem of physical interest, 
e.g. when two neutron stars coalesce.
At the present time such discussions
are beyond our means. Before trying to implement astrophysically relevant
initial data we must ensure that we can evolve the 
stellar perturbation equations accurately. Hence we have tested our 
evolution code by
experimenting with different kinds of initial data. This enables us to conclude
that our numerical implementation of the various equations is 
reliable. We have also gained some insight into the excitation of the stellar pulsation modes.

\subsection{Specifying initial data} 

To define acceptable initial data for the evolution problem is not
a trivial task. To specify astrophysically relevant initial
data one should first solve the fully nonlinear 3-dimensional
initial-value problem for (say) a newly formed neutron star that settles
down after core collapse. Then the results must be translated
into a form that makes them useful as initial data for the
perturbation equations. Although easy to describe in words,
each step is difficult and requires great care (cf. the analogous problem 
of two colliding black holes \cite{price94}). In short,
a detailed formulation of initial data for the 
perturbation equations
requires much further work, and we will return to it in the future.
Here we will focus on the evolution of the perturbation
equations from initial data that seems logical.

A careful analysis of the constraints that our initial 
data must satisfy indicates that we are free to choose 
our evolved variables $\{F,S,H\}$ such that the Hamiltonian
constraint and its first derivative are satisfied (of course
all boundary conditions and physical constraints must 
also be satisfied) \cite{init}. As an initial simplification we consider only time-symmetric initial data
($\dot{F}=\dot{S}=\dot{H}=0$).
This assumption leads to rather contrived initial data from 
a physical point of view. Basically,  the waveforms that we see as outgoing 
at future null infinity were initially incoming at past null infinity.
Nevertheless, time-symmetric initial data provide a useful starting point
for studies of the evolution problem. It should also be pointed out that
the time derivative of the Hamiltonian constraint $\dot{\cal H} = 0$ 
is automatically satisfied for time-symmetric data.

The choice of $\{F,S,H\}$ to satisfy ${\cal H}=0$ on the initial hypersurface
depends on the model problem under consideration. So far we 
have considered two different classes
of data sets. The first involves no initial fluid perturbation
in the star, while the second includes a nonzero fluid perturbation.

The first initial data set corresponds to the scattering of an 
incoming gravitational wave packet by the star. Here the fluid 
variable $H$ is set to zero, and the metric variables initially have support
only in the exterior vacuum. To enable comparison 
with the axial problem considered by Andersson and Kokkotas \cite{andersson96d}
we specify the Zerilli function, $Z$ to be a narrow Gaussian 
centered at a large radius. The spacetime variable $F$ can then be
calculated using
\begin{equation}
 F = r {\partial Z \over \partial r_\ast} + 
 { n(n+1)r^2 + 3nMr + 6M^2 \over r(nr+3M) } Z\ .				  
 \end{equation}
This equation follows when the definition (\ref{Zdef}) of the 
Zerilli function is combined with the constraint equation (\ref{Hamilton}).
Finally, we specify the remaining function $S$ by 
numerically integrating the constraint equation ${\cal H}=0$. 
An example of such initial data is shown in Figure~\ref{fig1}.

The scattering of gravitational waves is, however, not
the generic problem that we
are interested in. Generically, one would expect a non-vanishing 
perturbation inside the star. Thus, we would like to consider 
various perturbations in the stellar fluid. That is, we 
specify $H$ (arbitrarily) at some initial time. From the Hamiltonian 
constraint it is clearly seen that such a fluid perturbation must 
be accompanied by a non-vanishing metric perturbation. Thus, 
we should formally find $S$ and $F$ in such a way that the 
constraint equation (\ref{Hamilton}) is satisfied inside the star.
There is no a priori method of calculating $S$ and $F$ uniquely, since
we seemingly have the freedom to choose either of these two functions.
To test our numerical code, we have therefore made calculations for three 
different kinds of initial data: 
\begin{enumerate}
\item We combine the specified function $H$ with $F=0$. 
This is convenient since we then only need to integrate a first order 
differential equation (\ref{Hamilton}) to determine the corresponding $S$. 
\item An alternative to this is inspired by the weak-field limit.
It is known that (in the standard Regge-Wheeler
notation) $H_0-K \to 0$ as the star becomes more Newtonian \cite{thorne69}.
This would correspond to $S=0$. 
Once we have made this assumption we can numerically integrate the
second order
differential equation (\ref{Hamilton})
and find an acceptable $F$. 
\item  The third set of initial data is also based on the weak field
result. We set $F=2\delta U$, where
$\delta U$ is the perturbation in the Newtonian potential. Then
$\delta U$ satisfies  \cite{thorne69}
\begin{equation}
\nabla^2 \delta U = - 4 \pi \delta \rho = - {4\pi ( p+\rho) \over C_s^2 } H\ .
\end{equation}
Then the appropriate function $S$ is obtained by integrating (\ref{Hamilton}).
An example of this kind of initial data, for
\begin{equation}
H = C_s^2 (r/R)^2 \cos (\pi r/2R)   \ ,
\label{finit}\end{equation}
is shown in Figure~\ref{fig2}.
\end{enumerate}

\subsection{Numerical results}

\begin{figure}
\centerline{\epsfxsize=18cm \epsfbox{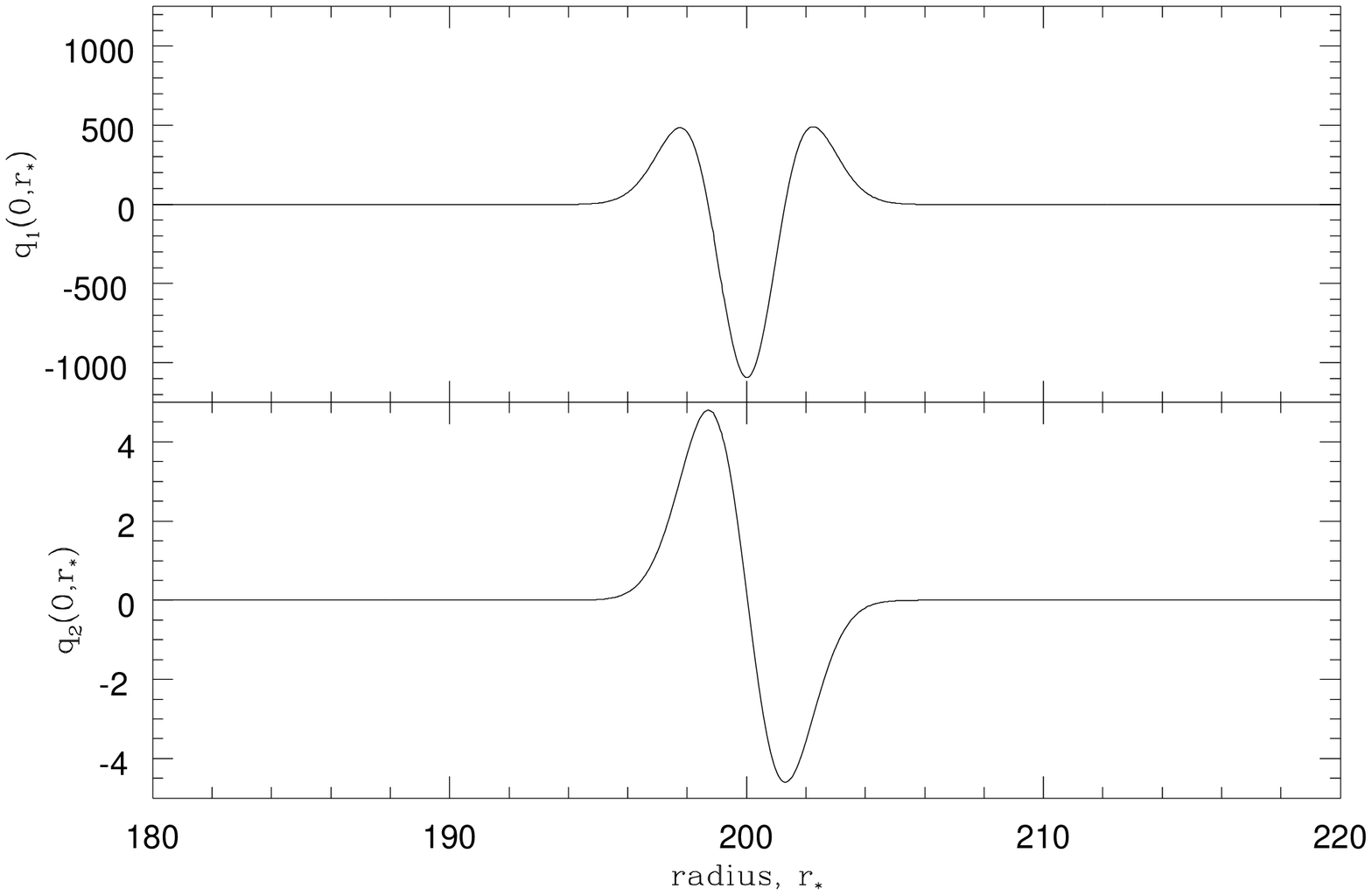}}
\caption{Initial data corresponding to a Gaussian pulse in the Zerilli
function. We show the two gauge-invariant quantities $q_1$ and $q_2$ 
(defined in section IIC).}
\label{fig1}\end{figure}

\begin{figure}
\centerline{\epsfxsize=18cm \epsfbox{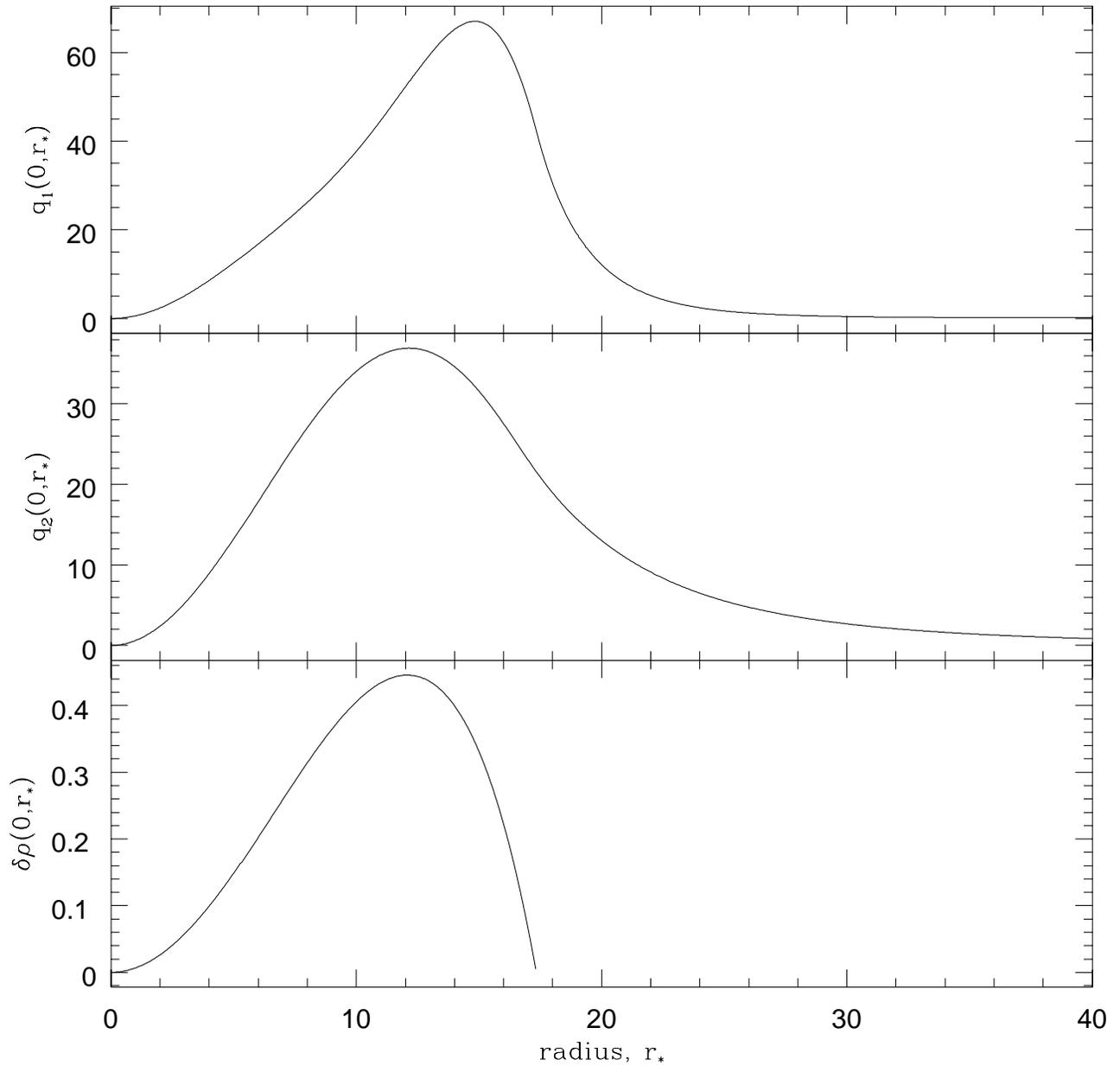}}
\caption{Initial data for the case when an initial perturbation 
in the fluid is specified. Here we use $H=A C_s^2 (r/R)^2 \cos (\pi r/2R)$.
and solve the constraint equations to obtain the required $S$ and $F$. 
These then lead to the 
two gauge-invariant quantities $q_1$ and $q_2$ 
(defined in section IIC) shown here.}
\label{fig2}\end{figure}

Even though the initial data for the model scenarios that we have chosen
(see Figures~\ref{fig1} and \ref{fig2}) are rather different, the 
gravitational waves that emerge from the system during each
evolution are qualitatively quite similar.

\begin{figure}
\centerline{\epsfxsize=16cm \epsfbox{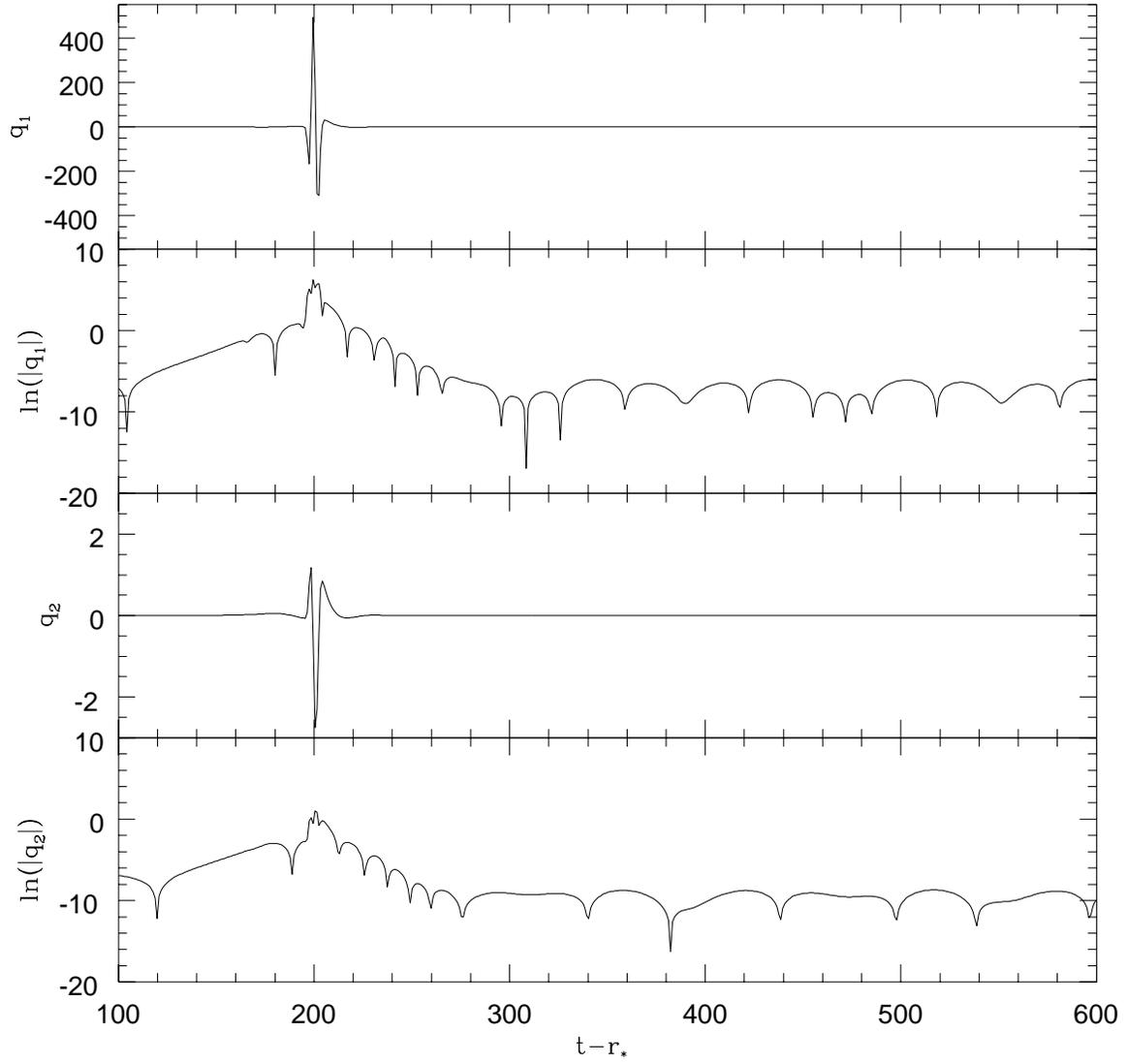}}
\caption{Evolution of the gauge-invariant quantities $q_1$ and $q_2$
for initial data corresponding to a Gaussian Zerilli function, cf. 
Fig. 1.}
\label{fig3}\end{figure}

In Figure \ref{fig3} we show the gravitational waves that follow when 
a Gaussian in the Zerilli function is scattered off the star. The
corresponding
initial data is shown in Figure \ref{fig1}. We use the gauge invariant
quantities
$q_1$ and $q_2$ (cf. (\ref{ginvar})) to represent the gravitational waves
that reach a distant observer. In this specific example, the observer is
located at $r_\ast = 200 M$.
After a sudden burst of waves follows a ringdown corresponding
to the quasinormal modes of the star. The ringdown  
consists of two parts. The first part (from $t-r_\ast\approx 200 - 280M$ in
Fig. \ref{fig3}) shows the high 
frequencies and the rapid damping
that are characteristics of the gravitational-wave 
modes. The second part of the signal (for $t-r_\ast > 280M$) is 
slowly damped, 
and the oscillations have longer wavelength. This part of the signal 
should correspond to the fluid
pulsation modes.
A spectral analysis of the signal shows that the 
emerging waves are composed of the $f$-mode, the first $p$-mode 
and the slowest damped $w$-mode for the star, cf. Figure \ref{fig5}. 
										  
\begin{figure}
\centerline{\epsfxsize=16cm \epsfbox{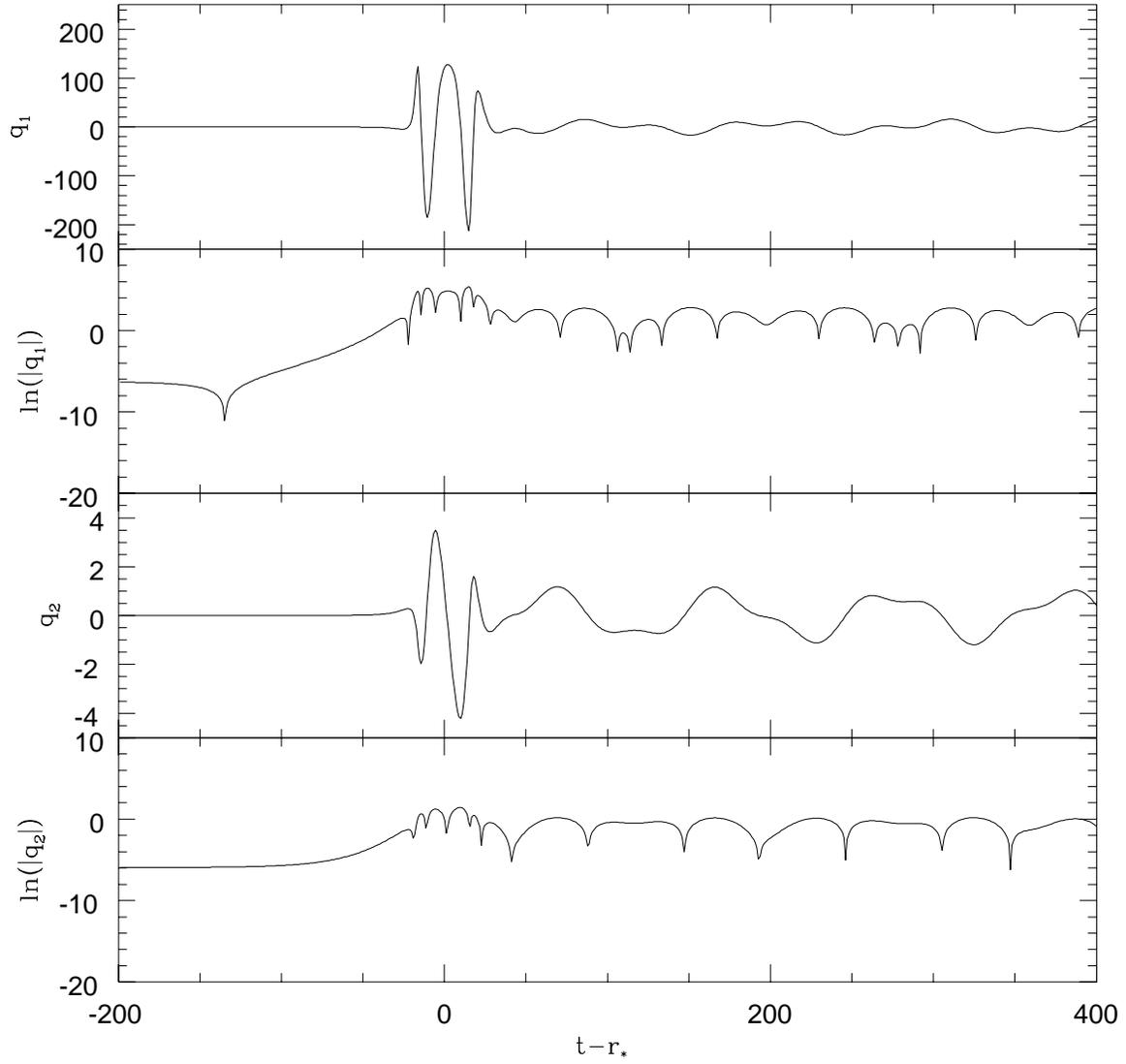}}
\caption{Evolution of the gauge-invariant quantities $q_1$ and $q_2$
for the initial data shown in Fig. 2.}
\label{fig4}\end{figure}

\begin{figure}
\centerline{\epsfxsize=14cm \epsfbox{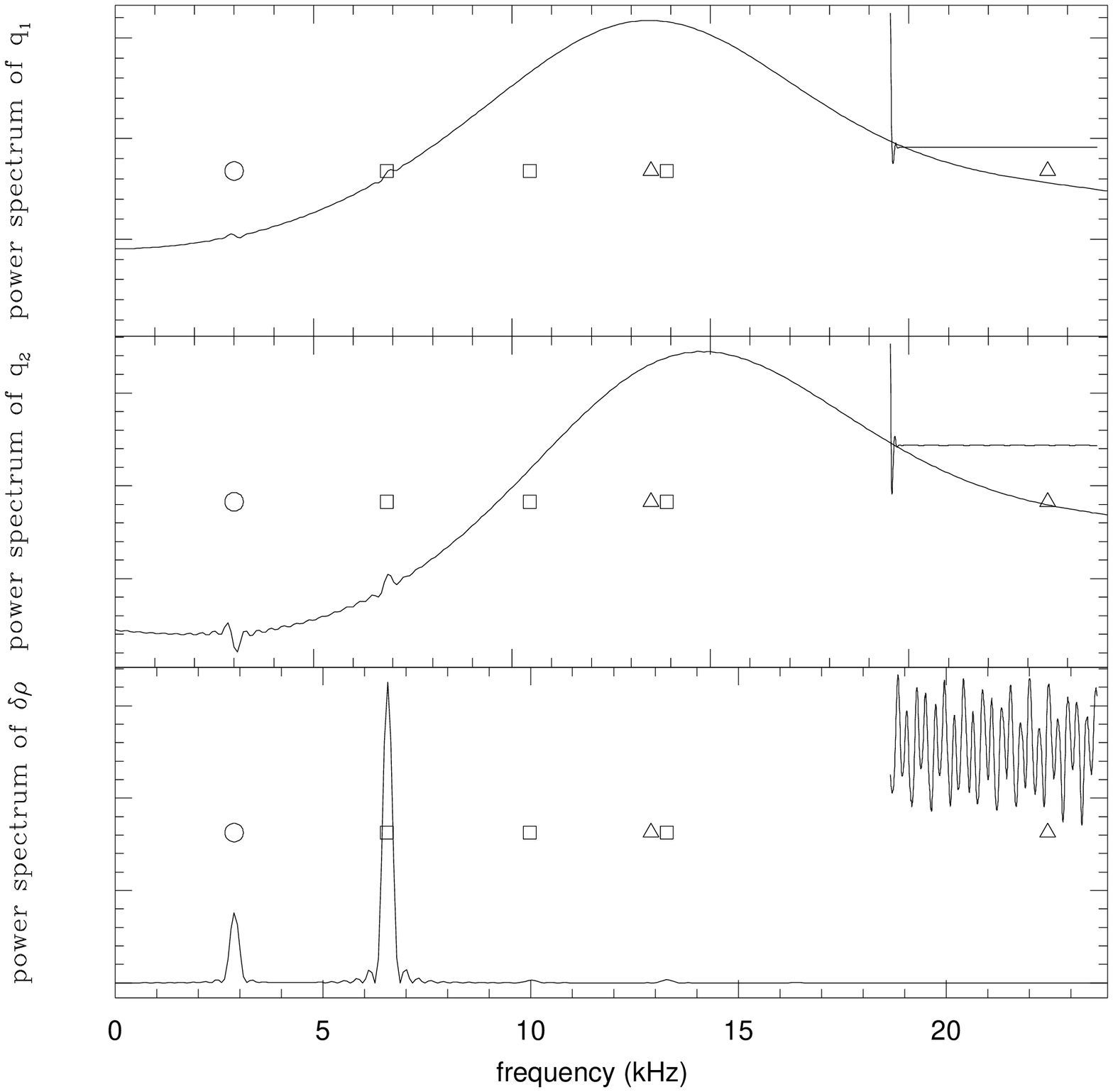}}
\caption{Power spectrum for the data in Fig. 3. The part of the 
signal that is used is shown in the upper right-hand corner of each panel.
The two upper panels show the power spectra for
the gauge-invariant quantities $q_1$ and $q_2$, respectively.
The lower panel shows the power spectrum for the fluid motion (as represented
by the Eulerian variation in the density $\delta \rho$).
In each panel we also show the position of the various pulsation modes of
the star. The $f$-mode is represented by a circle, the $p$-modes by 
squares, and the $w$-modes by triangles. It is clear that the emerging
gravitational waves contain the first $w$-mode together with the $f$-mode
and the first $p$-mode.}
\label{fig5}\end{figure}

\begin{figure}
\centerline{\epsfxsize=14cm \epsfbox{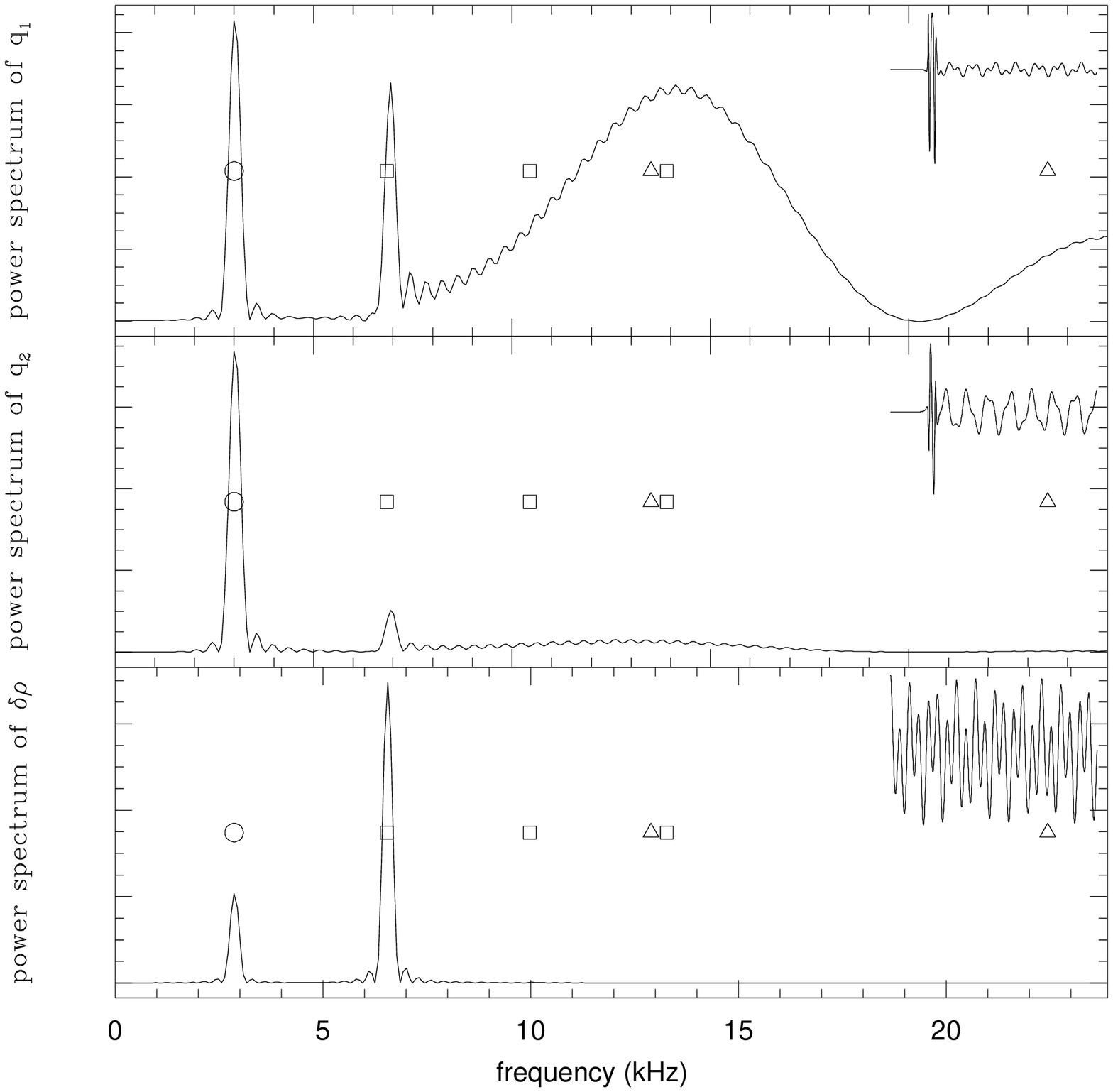}}
\caption{Power spectrum for the data in Fig. 4. The part of the 
signal that is used is shown in the upper right-hand corner of each panel.
The two upper panels show the power spectra for
the gauge-invariant quantities $q_1$ and $q_2$, respectively.
The lower panel shows the power spectrum for the fluid motion (as represented
by the Eulerian variation in the density $\delta \rho$).
In each panel we also show the position of the various pulsation modes of
the star. The $f$-mode is represented by a circle, the $p$-modes by 
squares, and the $w$-modes by triangles. It is clear that the emerging
gravitational waves contain the first $w$-mode together with the $f$-mode
and the first $p$-mode.}
\label{fig6}\end{figure}

The result is similar in the case when the stellar fluid
is perturbed initially. An example of such an evolution is shown in 
Figure \ref{fig4}. This example corresponds to the initial data (\ref{finit}),
cf. Figure \ref{fig2}.
Again, the post-burst signal can be roughly decomposed into two
parts: the first part ( $t-r_\ast\approx 0-25M$) corresponds to the
$w$-modes and the later signal ($t-r_\ast>25M$) is due to the 
fluid modes. The corresponding spectrum
shows  the presence of the $f$-mode, the first $p$-mode 
and the slowest damped $w$-mode, cf. Figure \ref{fig6}.  

These two examples show that one should expect 
both fluid and gravitational pulsation modes to be excited in a 
generic situation. But it is also clear that the relative amplitudes
of the various modes depend on the initial data. 
In the case of a gravitational-wave pulse impinging on the 
star (Figures~\ref{fig3} and \ref{fig5}) more energy is released 
through the
$w$-modes than in the case of an initially perturbed fluid
(Figures~\ref{fig4} and \ref{fig6}). This is not very surprising, since
in the first case the impinging waves must excite motion in the fluid,
which then leads to gravitational waves with (say) $f$-mode
characteristics. 
In the second
case the fluid is already perturbed so one would expect a more pronounced
excitation of the fluid modes. 

Our results indicate that there can be situations where a
considerably amount of energy
is released through the $w$-modes. Interestingly, we find that the
three ad hoc ways to specify initial data with an initial fluid perturbation
( see the previous section) lead
to different predictions for the energy released through $w$-modes.
Basically, we find that the excitation of the $w$-modes increases with the 
initial value of $S$. That is, if we set $S=0$ the evolutions
only show a glimmer of $w$-mode oscillations. In contrast, the $w$-modes
dominate the signal in cases
where $F=0$ initially. The answer to the question whether the $w$-modes will be
excited to a detectable level in an astrophysically relevant situation
 requires a more detailed study. But it is now clear that an 
affirmative answer 
requires a considerable initial energy in the spacetime function $S$. 

\subsection{A brief survey of the literature}

Since we have found that several of the stars pulsation modes
are excited in a general scenario, 
it is wortwhile to compare our results to previous numerical
neutron star simulations. 
The available simulations fall 
into two categories: Simulations of coalescing stars 
and studies of rotating core collapse. 

Almost uniquely, studies of coalescing neutron stars use Newtonian
hydrodynamics and extract the gravitational waves via the quadrupole 
formula. As already mentioned above, this approach cannot
account for the $w$-modes, but should be able to reveal
the presence of the fluid pulsation modes --- provided that these
 are excited to a significant level. A survey of the literature
reveals several indications that the pulsation modes
are present in the gravitational wave signals from coalescing stars. 
Waveforms obtained by Nakamura
and Oohara \cite{nakamura91} show clear mode presence 
(cf. Figs. 5,8 and 11 in \cite{nakamura91}). Their Figure 5 is
actually very similar to our Figure \ref{fig4}.
Ruffert et al. \cite{ruffert} have also obtained
gravitational wave signals from coalescing stars that show
late-time oscillations. Their waveforms
and spectra (cf. Fig 24 and 27 in \cite{ruffert}) show oscillations 
at frequencies between 1.5-2 kHz. 

Another interesting piece of evidence can be found in the
results of Zhuge et al. \cite{zhuge94,zhuge96}. They model
coalescing polytropic stars using SPH. In their waveforms
(cf. Fig. 6 in \cite{zhuge94}) one can possibly see oscillations 
in the late-time waveforms, but these oscillations are tiny. 
However, all the calculated spectra 
show a clear peak beyond the cut-off frequency that is 
associated with the actual coalescence (cf. Fig. 12 in \cite{zhuge94}). 
The authors attribute this peak to ``transient 
oscillations induced in the coalescing  stars during the merger
process --- the result of low-order $p$-modes with
frequencies somewhat higher than the Kepler frequency in the 
merging object". 

The situation is similar as far as rotating core collapse is concerned.
Most available studies use Newtonian hydrodynamics and account for
gravitational-wave emission through the quadrupole formula.
The collapse of a nonrotating star is expected to bounce
at nuclear densities, but if the star is rotating the collapse
can also bounce at subnuclear densities because of the
centrifugal force. In each case,  the emerging gravitational 
waves are dominated by a burst associated with the bounce. But 
the waves that follow a centrifugal bounce can also show
large amplitude oscillations that may be associated with
pulsations in the collapsed core.
Such results have been obtained by M\"onchmeyer et al. \cite{monch91}.
Some of their models show the presence of modes with different 
angular dependence superimposed
(see Fig. 5 in \cite{monch91}). Typically these oscillations have 
a period of a few ms and damp out in 
20ms. The calculations also show that the energy in the higher 
multipoles is roughly
three orders of magnitude smaller than that of the quadrupole.
More recent simulations by Yamada and Sato \cite{yamada} and
Zwerger and M\"uller \cite{zwerger} also show 
post-bounce oscillations (cf. waveforms in Figs. 2--4 of \cite{yamada}
and Figs. 5 and 6 of \cite{zwerger}).

There is thus evidence  that the fluid pulsation modes
of a star will be excited in both the collapse and the coalescence 
scenario \cite{liter}. From our present results it would then seem likely
that the $w$-modes will also be excited. But to show that this is
the case one must incorporate general relativity in the simulations
of collapse and coalescence. 
As yet there has been few attempts to do this, but an interesting
example is provided by the core-collapse
studies of Seidel et al. \cite{seidel88,seidel87,seidel90}. 
The approach adopted in those papers 
is, in fact, quite similar to our present one: One considers axial
(odd parity) or polar (even parity)
perturbations of a time-dependent background (that evolves
according to a specified collapse scenario). As expected
from the Newtonian studies, the extracted
gravitational waves are dominated by a sharp burst
associated with the bounce at nuclear densities. But there
are also features that may be related to the 
$w$-modes. In a case designed to collapse and rebound
with extreme energy Seidel et al. find a ringing mode in the
post-bounce gravitational waves (cf. Figure 7b of \cite{seidel87}).
Since there are no axial fluid modes for a nonrotating star, it
is plausible that this mode corresponds to one of the axial
$w$-modes of the core.  Furthermore, the power spectrum
for one of the simulations discussed in \cite{seidel88} (cf.
their Fig. 2)
shows some enhancement around
7 kHz. As the authors state, this ``is probably a result of some 
numerical noise...
rather than a true physical effect'', but it is 
interesting to note that the slowest damped axial $w$-mode has a 
pulsation frequency of roughly this magnitude.

\section{Concluding discussion}

In this paper we have discussed the evolution problem for the 
equations that govern a linear perturbation of a relativistic star.
We have taken the first steps towards developing a framework within which
one can study the behaviour of a neutron star at the late stages after a 
gravitational collapse or a binary coalescence. Such a framework would
provide a powerful testbed for fully relativistic simulations.
Furthermore, we can hope to learn much physics from the present approach.

The results we have presented indicate that
the gravitational waves that are generated by a perturbed neutron star will
carry the signature of both the gravitational-wave modes and the fluid
pulsation modes. This is a significant result because it indicates that
present estimates of the gravitational-wave emission (that are mainly based
on Newtonian calculations and the quadrupole formula) may be flawed.
Such studies would never yield the $w$-mode part of the signal, so the
energy that is released through these modes would not be accounted for.

However, to show that the $w$-modes are astrophysically relevant
(here we have only provided evidence that they can carry a considerable 
energy in certain model situations) we must consider the 
question of initial data more carefully. This can be done either by
extending our equations to the case of gravitational collapse 
(allowing the  background to be time dependent), or 
 by incorporating a $t=$constant data set from a fully 
relativistic 3D collapse. After extracting the perturbations
one can then  evolve them on a static background.
We will return to this problem in the near future.

Finally,  the present results show that it is meaningful to discuss whether
one can extract information about the parameters of the star from 
observed mode-signals. Specifically, since both the $f$-mode and the 
slowest damped $w$-mode are present in the signals in Figures~\ref{fig3}
and \ref{fig5} one could use the scheme that was suggested
by Andersson and Kokkotas \cite{andersson96d} to infer (or at least
constrain) the mass and the radius of the star.

\section*{Acknowledgements}

We acknowledge helpful discussions with Horst Beyer,  Curt Cutler, Lee Lindblom,  Philip Papadopoulous,
Bernd Schmidt and Ed Seidel.
This work was supported by an exchange program from the British Council 
and the Greek GSRT, and NA is supported by NFS (Grant no PHY 92-2290) 
and NASA (grant no NAGW 3874). NA and KDK also thank the Albert-Einstein
Institute in Potsdam for generous hospitality.

\section*{Appendix. The numerical evolutions}

We discretized the evolution equations (\ref{Swave}), (\ref{Gwave}) and 
(\ref{Hwave}) 
in a standard way, using second order 
centered finite differencing, on a 
regular grid. The grid is equally spaced in the time coordinate $t$ and
the spatial coordinate $r_\ast$. Since the coefficients in the 
equations are functions of the Schwarzschild radius $r$, we 
have to solve equation (\ref{tort}) numerically. 

Some of the coefficients in the evolution equations diverge at the origin,
and the finite difference equations cannot be used close to $r=0$. 
However, as discussed 
in section IID the dynamical variables all approach zero as $r\to 0$, 
and in practice it
was sufficient to use the leading order behaviour (\ref{centre}) 
close to the origin. 

The boundary equations (\ref{Hsurf}) and (\ref{Gsurf}) were used to evolve the functions $H$ and
$F$ at the stellar boundary, and $S$ was evolved using 
the evolution equation (\ref{Swave}).
A standard outgoing wave boundary condition was implemented for $S$ and $F$
at the outer
boundary of the grid. This causes some reflection from the
boundary, since in both cases the potential in the wave equation is small, 
but not zero. These reflections were too small to be ``seen'' in the
numerical data, but to be safe the outer boundary was placed 
sufficiently far away  
that it could have no influence on the results.

To construct initial data the Hamiltonian constraint (\ref{Hamilton})
must be solved. For the cases we considered this reduced to 
solving either a first order ODE for $S$ (when $F$ and $H$ were given)
or a second order ODE for $F$ (when $S$ and $H$ were given). 
The first order ODE was solved using a 
fourth order Runge-Kutta method \cite{press86}, and the second order ODE 
was solved using a relaxation method. A potential problem with finding
the solution at the stellar boundary (where 
the second space derivative of $F$ can be discontinuous if $H \neq 0$,
see Section IID)
was avoided by only using initial data with $H=0$ at the boundary.
 
Analytically, the Hamiltonian constraint ${\cal H}=0$ is satisfied 
throughout 
the evolution if it is satisfied initially. 
Numerically, there will always be some errors, and  
we can calculate  ${\cal H}$ in the discrete 
$L_2$ norm over the grid on each $t=$constant slice and
plot the result as a function of time. Calculating this error using 
different grid resolutions for a particular model
allows the convergence of the code to be tested. 
For all the results reported here second order 
convergence was found.

\pagebreak

\pagebreak

\pagebreak

\pagebreak

\pagebreak

\pagebreak

\begin{references}
\bibitem{angular} If the angular dependence of the perturbed quantities were to be accounted for
we should use
\begin{displaymath}
h_{\theta\theta}(t,r,\theta,\varphi)  =  r \sum_{\ell,m} F_{\ell m} (t,r) 
Y_{\ell m}
(\theta,\varphi) \ ,
\end{displaymath}
etcetera, where $Y_{\ell m}(\theta,\varphi)$ are the standard spherical 
harmonics.  The fluid variables
$\xi^i$, $\delta p$ and $\delta \rho$ are, of course, also expanded in 
spherical harmonics. For a more detailed description of the appropriate
angular functions, see  \cite{thorne67,kojima92}.
Our shorthand notation should not lead to any confusion since we focus 
on the perturbation functions and never actually reconstruct the perturbed
metric. Furthermore, it should be noted that
 we only consider the quadrupole case ($\ell=2$) in our numerical 
examples, and since the background geometry is spherically symmetric
it is sufficient to consider $m=0$.
\bibitem{origins} The wave equation for $S$ comes from the 
$[\theta-\varphi \ \theta-\varphi]$ component of the 
perturbed Einstein equations, the equation for $F$ 
follows when one combines the $[t\ t]$ and
the $[r\ r]$ components of the perturbed Einstein 
equations, and the wave equation for the variable $H$ follows from the 
time-derivative of the $t$-component of the equations of motion for the
fluid.
The Hamiltonian constraint corresponds to the 
$[t\ t]$-component of the perturbed Einstein equations.
For more details of the origin of all
these equations we refer to \cite{kojima92,kind93}.
\bibitem{init}
Appropriate initial data must satisfy the Hamiltonian constraint (\ref{Hamilton}),
together with the two momentum constraints (that follow from the
$[t\ r]$ and
the $[t\ \theta]$ components of the perturbed Einstein equations 
\cite{kojima92}). In order for the Hamiltonian constraint to be preserved
during the evolution we must also require that the first
time derivative of ${\cal H}$ vanishes. However, if we write the two momentum
constraints schematically as
\begin{eqnarray*}
{\cal M}_r (F,S,H,h_{tr},\xi^r) &=& 0 \ , \\
{\cal M}_\theta (F,S,H,h_{tr},\xi^r) &=& 0 \ ,
\end{eqnarray*} 
one can show that \cite{kind93}
\begin{displaymath}
\dot{\cal H} = {\cal M}_r^\prime + {\cal M}_r + {\cal M}_\theta \ .
\end{displaymath} 
Now, the Hamiltonian constraint and its first time derivative involves
only the variables which are evolved using equations (\ref{Swave}),
(\ref{Gwave}) and (\ref{Hwave}). 
Namely, the set $\{F,S,H,\dot{F},\dot{S},\dot{H}\}$. 
Thus we need only require that our initial data satisfies 
${\cal H} = \dot{\cal H} = 0$.
The two momentum constraints then
act as definitions of the metric 
perturbation $h_{tr}$ and the fluid displacement $\xi^r$.
\bibitem{liter} It is worth pointing out that although the various
results in the literature indicate that stellar pulsation modes
are present in the gravitational-wave signals at late times it is not
trivial to infer exactly which the relevant modes are. For example, 
in the case of core collapse a mode-analysis is complicated by 
the fact that the pulsating core is both rotating (most likely quite rapidly) 
and accreting matter. In comparison, our model evolutions are
very clean.
\end{references}
\end{document}